\documentclass[12pt]{elsart}
\usepackage{epsfig}
\usepackage{amssymb}

\renewcommand{\thefootnote}{\fnsymbol{footnote}}

\newcommand{\eqref}[1]{(\ref{#1})}

\begin{document}
\begin{frontmatter}

\title{Percolation approach to quark gluon plasma \\ in
high energy $pp$ collisions}
\author{J. Dias de Deus} and
\author{A. Rodrigues\thanksref{a}}
\address{CENTRA and Departamento de F\'{\i}sica (IST),\\
Av. Rovisco Pais, 1049-001 Lisboa, Portugal}
\thanks[a]{work at: Escola Secund\'{a}ria da Ramada, Portugal}
\begin{abstract}
We apply continuum percolation to proton-proton collisions 
and look for the possible threshold 
to phase transition from confined nuclear matter to quark gluon 
plasma. 
Making the assumption that J/$\psi$ suppression is a good signal
to the transition, we discuss this phenomenon for $pp$ collisions, in
the framework of a dual model with strings.
\end{abstract}
\end{frontmatter}

In recent years, high energy heavy ion collision experiments have been trying
to collect information on the possible existence of the plasma of quarks
and gluon (QGP).
One of the strategies has been to look for differences in particle production
between high density central heavy ion collisions, and low density ion 
collisions, nucleon-nucleus collisions and nucleon-nucleon collisions.
At the CENR/SPS ($\sqrt{s} \approx 19$ GeV) and now at 
Brookhaven/RHIC ($\sqrt{s} \approx 130-200$ GeV) several important 
general results have been obtained.

The charged particle density was found to increases with energy 
and the number of 
participating nucleons. The average transverse momentum also increases with 
energy and particle density, $<\!\!p_{T}\!\!>$ increases as well 
with the mass of the
produced particle. Strangeness increases with energy and particle 
density \cite{ref:1}. All these results, naturally excluding the 
dependence on the number of participants, are qualitatively similar to 
results obtained in nucleon-nucleon and nucleon-nucleus collisions 
\cite{ref:2}. None of them, separately, can then be taken as clear 
evidence for the formation of the QGP.

On the other hand, the anomalous suppression of the ratio J/$\psi$ over 
Drell-Yan production 
\cite{ref:3}, at large associated transverse energy $E_{T}$, has been
widely accepted as good signal for QGP formation \cite{ref:4}. In fact,
no such effect was seen in lower density nucleus-nucleus, nucleon-nucleus or
nucleon-nucleon collisions.

In this note, we shall argue that if the J/$\psi$ suppression has its origin in
the creation of an extended colour conducting region, as in percolation, the
same kind of suppression should occur even in nucleon-nucleon 
($pp$ or $p\bar{p}$) collisions at high enough energy.

We shall work here in the framework of multi-collision models, namely the 
Dual Parton Model (DPM) \cite{ref:5}, but try to be as general as 
possible. The basic ideas are: 1) nucleus-nucleus collisions can be built, in a
non trivial manner, from nucleon-nucleon collisions; 2) nucleon-nucleon 
collisions occur with formation of $2k$ intermediate strings, 
strings are always
in pairs, $k\geq1$ being a function of energy; and 3) strings may fuse and 
percolate \cite{ref:6} a process destroying naive additivity of elementary 
collisions.

The key parameter in transverse plane string percolation is the dimensionless
tranverse density $\eta$, with
\begin{equation}
  \eta = \frac{r_s^2}{R^2} N_s  \,,      \label{eq:eta}
\end{equation}

where $r$ is the tranverse radius of the string (we shall take $r=0.2$ fm),
$R$ the radius of the interaction area, and $N_{s}$ the number of
strings. Percolation occurs, in the $R\rightarrow\infty$, 
$N_{s}\rightarrow\infty$ limit, for $\eta\geq\eta_c \approx 1.15$.

As fusion and percolation of strings also occur in nucleon-nucleon 
collisions it is clear that the J/$\psi$ over D.Y. ratio will be strongly 
affected as $\eta$ approaches $\eta_c$. In the $pp$($p\bar{p}$) 
case~\eqref{eq:eta} becomes
\begin{equation}
  \eta = \left(\frac{r_s}{R_p}\right)^2 2k        \label{eq:eta1}
\end{equation}
where $R_p$ is the effective proton radius. As the observed increase of 
particle densities with energy is mostly due to the increase in the number
of formed strings, $k$ is also an increasing function of energy. Thus
$\eta$ may reach $\eta_c$ and become larger than $\eta_c$. This implies 
anomalous J/$\psi$ suppression.

Following previous work \cite{ref:6,ref:7} we treat fusion and percolation 
of strings as a two dimensional continuum percolation problem 
\cite{ref:8}. We performed computer simulation by throwing $N$ discs
(of radius $0.2$ fm) into a uniform region (of radius of order of the 
radius of the proton, $1$ fm) and counting the fraction of events $f(\eta)$
with percolation. In the $R\rightarrow\infty$, $N\rightarrow\infty$ limit
$f(\eta)$ becomes a step function with a sharp change at $\eta=\eta_c$. As
$r/R$ is not so small, finite size effects are important, affecting mostly the
slope ($a$) of the function at $\eta_c$, but the value of $\eta_c$ itself.
The computer simulation results were fitted by the function,
\begin{equation}
  f(\eta) = \left( 1 + e^{-(\eta - \eta_c)/a} \right)^{-1}
          \label{eq:fperc}
\end{equation}
and the following values were found for the parameters: 
$a=0.1666\pm0.0067$ and $\eta_c=1.3584\pm0.0116$ (see Fig. 1)

\begin{figure}
  \begin{center}
  \mbox{\epsfig{file=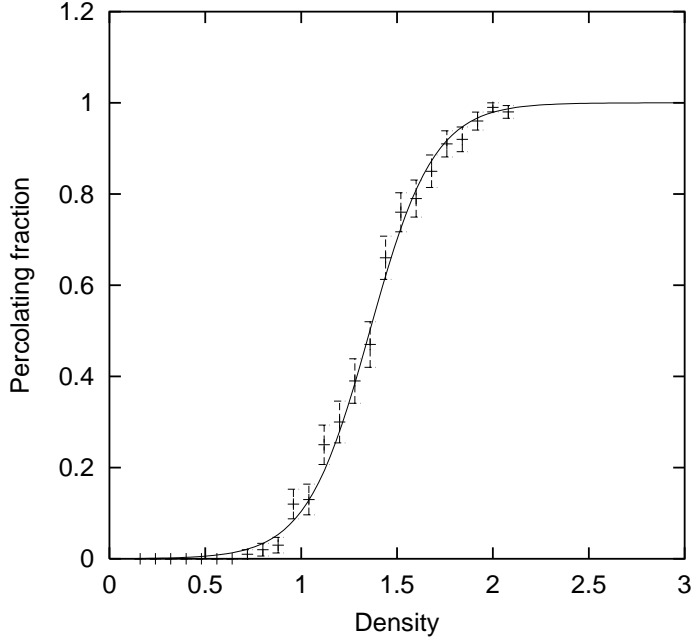,width=0.7\textwidth}}
  \end{center}
  \caption[percolating fraction]{Percolation probability as function of 
  the transverse dimensionless density $\eta$ for pp collisions geometry.
  The curve shows the fit with~\eqref{eq:fperc}. We obtain 
  $a=0.1666\pm0.0067$ and $\eta_c=1.3584\pm0.0116$.}\label{fig:1}
\end{figure}

Assuming now, as in \cite{ref:7}, that the J/$\psi$ production is 
prevented in a plasma of colour charges \cite{ref:4} and that such 
situation corresponds to percolation and creation of a large conducting area
\cite{ref:6} we obtain for the J/$\psi$ over Drell-Yan ratio,
\begin{equation}
  R(\eta) = K(1-f(\eta)) = K\left[ e^{(\eta - \eta_c)/a} + 1 \right]^{-1} \,,
          \label{eq:4}
\end{equation}
where $K\approx55$ \cite{ref:9} is the value of the ratio at 
moderate energy.

The problem now is simply the problem of relating $\eta$ 
to $\sqrt{s}$,~\eqref{eq:eta1}. 
In other words, we need to obtain a reasonable estimate
for the energy dependence of $k$. This is what we shall attempt now.

If in nucleus-nucleus collisions $<\!\!\nu\!\!>$ is the average number of
nucleon-nucleon collisions and $2k$ is the number of strings per 
nucleon-nucleon collision, the average number of strings $N_s$ is
given by,
\begin{equation}
  N_s = <\!\!\nu\!\!>2k\,.
          \label{eq:5}
\end{equation}

In nucleon-nucleon collisions $<\!\!\nu\!\!>=1$ and $N_s=2k$. 
From~\eqref{eq:eta} and~\eqref{eq:5} the condition 
for the percolation transition, in the case of $r/R\ll 1$, is
\begin{equation}
  \eta = \eta_c =\left(\frac{r}{R}\right)^{2}<\!\!\nu\!\!>2k\approx1.15\,.
          \label{eq:6}
\end{equation}

By interpreting the NA50 anomalous J/$\psi$ suppression at $\sqrt{s}\approx
19$ GeV as the result of percolation, one can try to estimate the 
number of formed strings in nucleon-nucleon collisions. The basic 
information is that anomalous
suppression is absent in S-U central collisions, but it is present in Pb-Pb
central collisions \cite{ref:3}. This means, $\eta_{S-U}<1,15$ and
$\eta_{Pb-Pb}>1,15$.

At relatively low energy, it is known, from NA49 and WA98, SPS experiments,
that the number of nucleon-nucleon collisions in central nucleus-nucleus
collisions is, roughly,(see, for instance \cite{ref:11}).
\begin{equation}
<\!\!\nu\!\!>\backsimeq1.5\, \frac{N_p}{2}  \,,
          \label{eq:7}
\end{equation}
where $N_p$ is the number of participants in a central AB, A$\leq$B, 
collision,
\begin{equation}
  N_p\backsimeq A^{^{\frac{2}{3}}}(A^{^{\frac{1}{3}}}+B^{^{\frac{1}{3}}})\,.
          \label{eq:8}
\end{equation}
On the other hand
\begin{equation}
  R\backsimeq A^{^{\frac{1}{3}}} fm\,.
          \label{eq:9}
\end{equation}
By using~\eqref{eq:7},~\eqref{eq:8} and~\eqref{eq:9} in~\eqref{eq:6} we
obtain, from S-U
\begin{equation}
   k(\sqrt{s}\backsimeq19) \lesssim 1.7\,,
          \label{eq:10}
\end{equation}
and, from Pb-Pb,
\begin{equation}
   k(\sqrt{s}\backsimeq19) \gtrsim 1.6\,.
          \label{eq:11}
\end{equation}
As we are not so confident with these estimates we shall include an error 
of the order of 15\% and study the ratio J/$\psi$ over DY in the range
\begin{equation}
 1.4 \lesssim k(\sqrt{s}\backsimeq19) \lesssim 1.9\,.
          \label{eq:12}
\end{equation}

If we look now at the charged particle densities in $pp$($p\bar{p}$) 
collisions, in
the spirit of DPM, we have that particles are emitted from two kinds of 
strings: valence strings(V), always 2 from valence quark interactions and 
shorter sea strings(S), from sea parton interactions, in a number growing with
energy, $2(k-1)$. The central charged particle density is written as(see, for
instance, \cite{ref:10} and \cite{ref:11}),
\begin{equation}
   \frac{dN}{dy}\Bigg\vert_{pp}=2\frac{dN}{dy}\Bigg\vert_{V} + 2(k-1)
\frac{dN}{dy}\Bigg\vert_{S}\,.
          \label{eq:13}
\end{equation}
On the right hand side of~\eqref{eq:13} we have both contributions, from 
V and S strings.

We assume that $\frac{dN}{dy}\vert_{V}$ and 
$\frac{dN}{dy}\vert_{S}$ are constant ("Feynman scaling") and that 
the observed rise of the plateau is determined by the increase in the 
number of strings, i.e., by increase of $k$. In the low energy limit, 
$k\rightarrow 1$, we thus have Feynman scaling with, from data 
\cite{ref:11,ref:12},
\begin{equation}
   \frac{dN}{dy}\Bigg\vert_{pp} \overrightarrow{k\rightarrow 1} \,\,\,
 2\frac{dN}{dy}\Bigg\vert_{V} \backsimeq 1.45\pm 0.05
          \label{eq:14}
\end{equation}

\begin{figure}
  \begin{center}
  \mbox{\epsfig{file=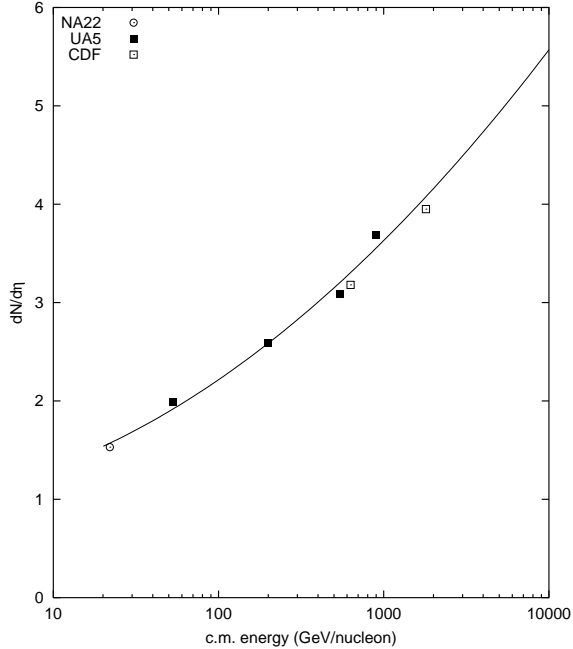,width=0.55\textwidth}}
  \end{center}
  \caption[Data from pp]{Pseudo-rapidity density as function of 
  c.m. energy. The solid line
  represents parameterization used to fit the data. 
  }\label{fig:2}
\end{figure}

In order to determine the energy dependence of $k$ we do the following.
Solve de equation~\eqref{eq:13} for $k$, with $\frac{dN}{dy}\vert_{V}$
fixed by~\eqref{eq:14} and using for $\frac{dN}{dy}\vert_{pp}$ a 
parameterization to the $pp$($p\bar{p}$) data \cite{ref:11},
\begin{equation}
   \frac{dN}{dy}\Bigg\vert_{pp} = 0.957+0.0458\ln{\sqrt{s}}+
   0.0494\ln^{2}{\sqrt{s}}
          \label{eq:15}
\end{equation}
 
For each of the limiting values of $k(\sqrt{s}\approx 19)$~\eqref{eq:12}
, combined with the limiting values 
of $\frac{dN}{dy}\vert_{V}$~\eqref{eq:14}, we adjust the constant
$\frac{dN}{dy}\vert_{S}$ to obtain agreement with~\eqref{eq:15}.
The fit to $\frac{dN}{dy}\vert_{pp}$ high energy 
data for $\sqrt{s}\geq 20$ GeV is shown in (Fig.~\ref{fig:2}). 

\begin{figure}
  \begin{center}
  \mbox{\epsfig{file=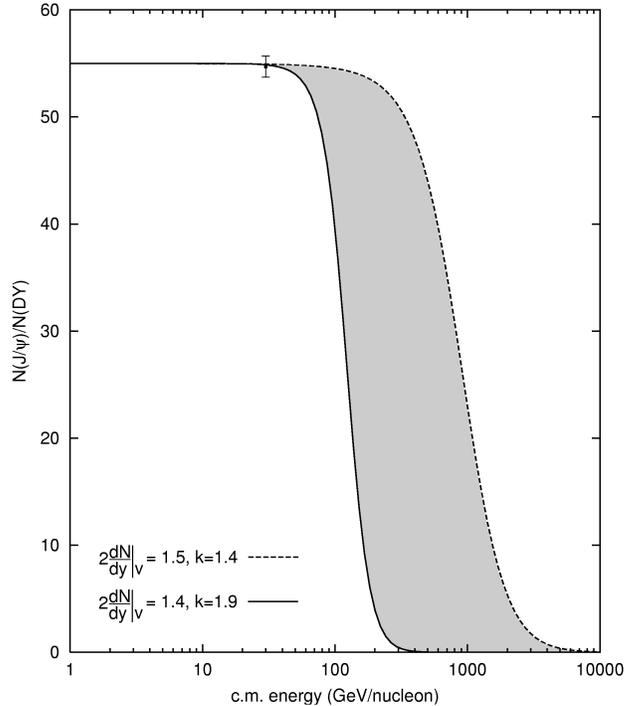,width=0.6\textwidth}}
  \end{center}
  \caption[jpsi]{The ratio of J/$\psi$ to Drell-Yan events as
  function of c.m. energy in $pp$ collisions.}\label{fig:3}
\end{figure}

In Fig.~\ref{fig:3} we present the $\sqrt{s}$ dependence of the J/$\psi$
over DY ratio, equation~\eqref{eq:4} and~\eqref{eq:eta1}, for the two limiting
values of $k$. In conclusion: we expect a fast drop of J/$\psi$ over DY ratio
in $pp$($p\bar{p}$) collisions in the energy range 
$150 \lesssim \sqrt{s} \lesssim 1000$ GeV/nucleon. These are energies of RIHC
and Tevatron.

Our work can be criticized from several different points of view:
\begin{enumerate}
\item[1)] The J/$\psi$ over DY ratio is also affected by internal absorption 
and as the number of strings increases with energy, absorption makes the ratio 
continuously decrease with energy. This effect was not included.

\item[2)] As charm production probability increases with energy the 
J/$\psi$ over DY ratio should have a tendency to increase with energy. This
correction was also not included. One should, perhaps, in future consider 
the ratio J/$\psi$ over $c\bar{c}$ production as the reference quantity.

\item[3)] As in $pp$($p\bar{p}$) collisions the interaction radius 
increases slowly
with energy the critical value of $\eta$, $\eta_c$, should decrease with 
energy. However the important effect is the actual decrease of 
$\eta\sim\frac{1}{R^2}$ with the consequence that the percolation transition 
occurs at higher value of energy.

\item[4)] The model is purely soft model and hard effects related to the
  increase of $<\!\!p_{T}\!\!>$ and changes in multiplicities were 
not included. These effects can be accounted for with fusion of 
strings but were not considered here. They are being studied now.

\item[5)] Non-uniform distributions in impact parameter (like gaussian 
distributions) give rise to an increase of $\eta_c$ (see last paper 
in\cite{ref:6}), and consequently the percolation transition will 
tends to be displaced to higher energy.

\item[6)] One may question the validity of continuum percolation arguments 
when the ratio $r/R$ is so large, $r/R\backsimeq1/5$. Technically there is no 
problem, but we are not sure of the validity of the treatment.
\end{enumerate}

While finishing this paper we became aware of the work of T. Alexopoulos et al.
\cite{ref:13} dealing with evidence for deconfinement at Tevatron 
($\sqrt{s} =1.8$ TeV).

\section*{Acknowledgments}
We would like to thank Roberto Ugoccioni for help at several stages in 
this research project.



\end{document}